# Application of Shaken Lattice Interferometry Based Sensors to Space Navigation

Margaret M. Rybak[1], Penina Axelrad[2], Catie LeDesma[3], and Dana Z. Anderson[4],
*University of Colorado, Boulder, Colorado, 80309, USA*

and

Todd Ely[5]
*Jet Propulsion Laboratory, Pasadena, California, 91109, USA*

**Abstract**

High-sensitivity shaken lattice interferometry (SLI) based sensors have the potential to provide deep space missions with the ability to precisely measure non-gravitational perturbing forces. This work considers the simulation of the OSIRIS-REx mission navigation in the vicinity of Bennu with the addition of measurements from onboard SLI-based accelerometers. The simulation is performed in the Jet Propulsion Laboratory's (JPL) Mission Analysis, Operations and Navigation Toolkit (MONTE) and incorporates OSIRIS-REx reconstructed trajectory and attitude data from the Navigation and Ancillary Information Facility (NAIF) database. The use of the reconstructed data from NAIF provides realistic true dynamical errors and JPL's MONTE software allows for a high-fidelity simulation of a nominal reference for the filter. The navigation performance and reduction of tracking and complex modelling enabled by the onboard SLI-based sensor are presented for two orbital phases of the OSIRIS-REx mission. Overall, the results show that the addition of SLI-based accelerometer measurements improves navigation performance, when compared to a radiometric tracking only configuration. In addition, results demonstrate that highly-precise accelerometer measurements can effectively replace at least one day of DSN passes over a three-day period, thereby reducing tracking requirements. Furthermore, it is shown that lower-fidelity surface force modeling and parameter estimation is required when using onboard SLI-based accelerometers.

---

[1] Graduate Student, Smead Aerospace Engineering Sciences, University of Colorado, Boulder, margaret.rybak@colorado.edu.
[2] Professor, Smead Aerospace Engineering Sciences, Fellow AIAA, University of Colorado, Boulder, penina.axelrad@colorado.edu.
[3] Graduate Student, JILA Physics, University of Colorado, Boulder, catherine.ledesma@colorado.edu
[4] Professor, JILA Physics Fellow, University of Colorado, Boulder, dana@jila.colorado.edu
[5] Principal Investigator, Deep Space Atomic Clock, Jet Propulsion Laboratory, California Institute of Technology, todd.a.ely@jpl.nasa.gov.

## 1  Introduction

The design of trajectories for successful space exploration missions relies on detailed predictions of the gravitational and non-gravitational forces that will affect the spacecraft motion. During the mission, navigators continue to refine the trajectory predictions based on tracking data. The refinements include improving the fidelity of spacecraft surface models affecting solar radiation pressure and gravity models of target bodies being explored. We are interested in exploring how accelerometers of sufficient sensitivity at low frequencies could enable more precise characterization of non-gravitational forces acting on the spacecraft. This configuration could alleviate the modeling and tracking requirements for a mission, as well as provide better navigation performance. Advances in cold atom interferometry (CAI) have demonstrated the potential to provide highly sensitive measurements of acceleration with small biases, and negligible scale factor and drift (Jekeli, 2006; Battelier, et al., 2016; Reguzzoni, Migliaccio, & Batsukh, 2021). Shaken Lattice Interferometry (SLI) has further adapted CAI technology by trapping the atoms in the nodes of an electromagnetic standing wave. This configuration allows for longer interferometric sequence times and thus higher sensitivity than free-space propagation CAI-based sensors. The interferometric time for SLI-based sensors is not tied to the size of the device, as in the case of conventional CAI-based sensors, thereby supporting their potential miniaturization to dimensions suitable for use onboard spacecraft (Weidner, 2019; Theurkauf, 2020). This research explores how onboard measurements from SLI-based accelerometers could be leveraged in a deep space mission to reduce the modeling and tracking burden.

Ultra-sensitive electrostatic accelerometers (EAs) were used on both the Gravity Recovery and Climate Experiment (GRACE) and Gravity field and steady-state Ocean Circulation Explorer (GOCE) missions. The goal of these missions was to precisely map the Earth's gravity field using highly-sensitive onboard EAs to measure the gravity gradient and non-gravitational accelerations. The sensitive EAs were critical to the success of GRACE and GOCE, however they required frequent calibration due to their long term noise instability (Visser, 2008; Rummel, Yi, & Stummer, 2011; Christophe B. , 2013; Klinger & Mayer-Gurr, 2016; Peidou & Pagiatakis, 2019; Fan, Liang, Wang, & Luo, 2022). CAI-based accelerometers have been proposed for use on future gravity mapping missions due to their noise stability at low frequencies compared to the performance of EAs (Carraz, Siemes, Massotti, Haagmans, & Silvestrin, 2014; Chiow, Williams, & Yu, 2015; Hogan & Kasevich, 2016; Migliaccio, et al., 2019; Leveque, et al., 2021). Research has also considered the concept of hybridizing electrostatic and CAI-based accelerometers to capture the different sensor performance advantages (Jekeli, 2005; Christophe, et al., 2017).

Conventional and CAI-based accelerometers have also been considered for use in navigation applications. Prior work by Bhatia and Geller (2020) investigated the use of pairs of CAI-based accelerometers as ultra-precise inertial gradiometers for autonomous navigation. They evaluated the navigation performance of using measurements from onboard CAI-based accelerometers correlated to onboard gravity maps to determine orbital information. Canciani (2012) performed a simulation analysis of an aircraft using a combination of an aviation grade inertial navigation system (INS) and CAI-based accelerometer measurements to improve the INS performance. Jekeli (2005) presented analysis of the integration of CAI-based sensors into a free-inertial navigation system and implications for navigation error improvement. Both Canciani and Jekeli focused on the method of dead-reckoning, which uses accelerometer measurements to directly propagate the position and velocity states of a vehicle. Alternately, Ely, Heyne, and Riedel (2012) investigated a method for using measurements from a conventional accelerometer directly in a navigation filter for the Altair lunar lander.

Our work focuses on how an onboard SLI-based accelerometers could be leveraged to precisely measure the non-gravitational forces during a deep space mission. The method follows the approach used by Ely, Heyne, and Riedel where measurements from the accelerometer are used directly in the navigation filter. The Origins, Spectral Interpretation, Resource Identification, Security-Regolith Explorer (OSIRIS-REx/OREx) mission to the asteroid Bennu was selected as a case study for this research. Actual OREx mission data from the Navigation and Ancillary Information Facility (NAIF) is used to provide the most realistic dynamic error profiles possible. The acceleration magnitudes and profiles experienced onboard OREx are investigated to identify the performance metrics required for a viable SLI-based accelerometer. Simulated SLI-based accelerometer measurements are then integrated into a high-fidelity simulation, in the Jet Propulsion Laboratory's (JPL) Mission Analysis, Operations and Navigation Toolkit (MONTE), of the OREx navigation for two orbital phases. During the OREx mission, optical-navigation and landmark tracking were used in addition to DSN tracking. To understand the impact specific to using onboard accelerometers, this study considers estimation using only DSN tracking and SLI-based accelerometer measurements. The navigation performance is analyzed to determine the potential improvements in orbit estimation errors and reduction of tracking and modeling enabled by onboard SLI-based accelerometers.

The rest of this paper is organized as follows. Section 2 provides the background and current performance of CAI/SLI-based sensors and their advantages over current state-of-the-art EA technology. Section 3 describes the motivation for the selection of the OREx mission as a case study and details the dynamic models used to simulate the OREx orbit. Section 4 presents the methods for simulation of DSN and SLI-based accelerometer measurements and the navigation filter set-up. Section 5 provides an analysis of the navigation performance for the OREx mission simulation, using measurements from the DSN in addition to the onboard SLI-based measurements. The paper concludes in Section 6 with a summary of findings and suggestions for future work.

## 2    Cold Atom Interferometry Based Accelerometers

Conventional accelerometers, such as mechanical or electrostatic sensors, measure the restoring force required to hold a proof mass steady in a null position within the housing or platform on which the unit is mounted. The restoring force is sensed to determine the acceleration on the proof mass (Conklin, 2015). Electrostatic accelerometers (EA) were used onboard the GRACE and GOCE missions to sense non-gravitational forces, including drag, solar radiation pressure, Earth albedo, thruster firings and magnetic torque accelerations on the spacecraft. Accurate measurement of the non-gravitational forces combined with precise radiometric tracking during the GRACE and GOCE missions enabled high-fidelity resolution mapping of the Earth's gravity field (Touboul, Willemenot, Foulon, & Josselin, 1999). The EAs onboard the GRACE and GOCE missions achieved accuracies of $10^{-10}$ m/s$^2$/√Hz and $10^{-12}$ m/s$^2$/√Hz (normalized for a 1-s integration period) respectively (Touboul P. , 2003; Flury, Bettadpur, & Tapley, 2008). However, these sensitivities were restricted to specific measurement bandwidths, $3.5 \times 10^{-2}$ to 0.2 Hz for GRACE and $5 \times 10^{-3}$ to 0.1 Hz for GOCE. At lower frequencies, beyond these measurement bands, the noise increases as 1/f and with bias instability noise, typical for conventional accelerometers (Christophe, Marque, & Foulon, 2010; Christophe B. , 2013; Christophe, et al., 2017; Fan, Liang, Wang, & Luo, 2022). While this was not detrimental for the mission objectives, it does mean that the measurements could not be used directly, for example to support navigation (Rummel, Yi, & Stummer, 2011).

CAI-based accelerometers were first developed in the early 1990's and have been shown to provide sensitivities that far exceed the performance of conventional mechanical and electrostatic sensors (Clauser, 1988; Kasevich & Chu, 1991; Carnal & Mlynek, 1991). The interferometry sequence is performed by loading an atom cloud into a magneto-optic trap (MOT) where the atoms are cooled to micro-Kelvin level to reduce their velocities. The atom cloud is then excited by laser pulses, separating it into counter-propagating clouds along the axis of acceleration. The incident acceleration on the sensor causes a phase difference to accumulate between the two propagating clouds over the interrogation time. A final laser pulse recombines the clouds, generating an interference pattern representative of the difference in propagation paths. The acceleration can then be deduced from the interference pattern (Kasevich & Chu, 1991). The potential advantage of using CAI-based accelerometers is that, unlike EAs, the underlying physics does not introduce any systematic errors or drift, dramatically reducing the requirements for calibration. CAI-based sensors display white noise at all frequencies below at least 100 mHz (Canuel, et al., 2006; Meunier, et al., 2014; Dutta, et al., 2016). This facilitates their use in making absolute acceleration measurements which can be used directly for navigation (Canuel, et al., 2006; Christophe, et al., 2017; Abrykosov, et al., 2019; Trimeche, et al., 2019).

Longer interrogation times improve the phase difference resolution for CAI, thereby providing greater sensor sensitivity. On Earth, interrogation times of conventional CAI-based sensors are limited by the atom free-fall-time due to gravity. Ground based CAI-based accelerometer studies have shown potential acceleration sensitivities of $1\times10^{-8}$ m/s$^2$/$\sqrt{Hz}$ (Jekeli, 2005; Kasevich, 2007). However, in zero-g, much longer interrogation times can be achieved. Nyman et al. (2006) presented a potential sensitivity of $1.8\times10^{-10}$ m/s$^2$/$\sqrt{Hz}$ for a CAI-based accelerometer tested in zero-g, detailed in Table 1. The amplitude spectral power density is scaled for the 15 second shot time, as shown in Eq. (1).

Table 1: Nyman Accelerometer Best Case Performance Metrics (Nyman, et al., 2006)

| Amplitude Spectral Density | $1.8\times10^{-10}$ m/s$^2$ /$\sqrt{Hz}$ |
|---|---|
| Sensitivity per Shot | $4.7\times10^{-11}$ m/s$^2$ /shot |
| Total Time | 15 seconds |
| Interrogation Time | 10 seconds |

$$\sigma(15\ s) = \frac{1.8\times10^{-10} \frac{m}{s^2}}{\sqrt{Hz}\sqrt{15\ s}} = 4.7\times10^{-11} \frac{m}{s^2} \tag{1}$$

Unlike the continuous measurement capability of conventional accelerometers, CAI-based accelerometers can only measure acceleration during the interrogation time of each duty cycle. There are dead bands between measurements, during which the atom cloud is cooled and prepared for the next interferometry sequence (total time minus interrogation time, 5 seconds in the case in Table 1). Hybrid accelerometer configurations have been proposed, which use measurements from conventional EAs to fill in the CAI dead bands (Lautier, et al., 2014; Christophe, et al., 2017; Abrykosov, et al., 2019). Following the work presented in Jekeli (2005), measurements from the combination of multiple accelerometers can be treated as a single measurement with combined accelerometer noise properties. Meunier et al. (2014) proposed an alternative approach of joint interrogation of up to five different atom clouds to remove dead time; which also resulted in improved sensitivity of $1 \times 10^{-13}$ m/s$^2$ at a 60 s interrogation time. For the purposes of this work, it is assumed that either multiple CAI-based accelerometers are used with staggered sequence start times to cover dead bands, or the multi-cloud approach is used, thereby providing continuous measurement capability.

In most CAI-based sensors the cloud of cold atoms is released, and the necessary interferometer steps are performed during its free-fall. With this method, interrogation time is tied to the free-fall distance, and therefore the sensitivity is constrained by the size of the apparatus. The sensitivity can be decoupled from cavity size by instead, confining the atoms to an optical lattice, where splitting/recombination and propagation are carried out by modulating the phase of the lattice. Machine learning techniques are used to determine effective shaking functions to achieve the desired sensor response to input accelerations (Weidner, Yu, Kosloff, & Anderson, 2017; Weidner & Anderson, 2018; Weidner & Anderson, 2018; Weidner, 2019; Theurkauf, 2020). Using shaken lattice interferometry (SLI), improved sensitivities can be achieved by simply increasing the atom hold time after splitting. Hold times of up to 20 s, a duration completely infeasible in 1-g for conventional CAI, have already been realized experimentally (Xu, et al., 2019).

By confining atoms to a 3D lattice and loading atoms that have condensed into a Bose-Einstein Condensate (BEC), high atom numbers with narrow velocity distributions can also yield increased sensitivities. Atom-atom interactions, which occur when atoms become denser with decreasing temperature, can cause phase errors due to their effect on both atom coherence and heating. By implementing a 3D lattice, interactions can be lowered by sparse population of the lattice through the use of a deep two-dimensional lattice with low single-site occupation. This results in an array of 1-D low atom number interferometers with shaking then taking place along the third dimension. This approach allows total atom numbers to reach $10^6$, lowering shot noise to levels comparable with state-of-the art atom interferometers.

High cycling rates are also feasible for SLI if an all-optical approach is taken for the generation of BEC. This involves using high powered laser beams instead of magnetic fields to perform evaporative cooling where atoms are cooled to degeneracy and form a BEC. By dynamically controlling these optical traps, experimental cycle times under 2 s have been achieved (Albers, 2020) and those under 1 s proposed (Roy, Green, Bowler, & Gupta, 2016). SLI therefore has the potential to enable sensors with high sensitivities that are much smaller than existing CAI-based devices, more suitable for use on spacecraft.

## 3 OREx Case Study Simulation

OREx was selected as a case study for this research because it presents the unique navigation challenge of orbiting a low-gravity asteroid. Bennu's weak gravitational pull made OREx highly susceptible to non-gravitational perturbing forces, such as solar radiation pressure (SRP) and spacecraft thermal radiation. The success of the mission was achieved with a combination of extensive Deep Space Network (DSN) tracking, altimetry measurements, optical navigation, and the development of high-fidelity non-gravitational force models (Hesar, Scheeres, & McMahon, 2016; Williams, et al., 2018). The dominant non-gravitational perturbing force on the nominal OREx trajectory was SRP, and basic modeling proved insufficient in representing its full effect on the spacecraft. In certain orbit configurations, the SRP was on the order of Bennu's non-point mass gravity field perturbations, causing the errors in the SRP model to alias into the gravity field estimation. To reduce the navigation errors due to SRP and improve the gravity recovery, the navigation team created a complex ray-trace SRP model. This model included self-shadowing and effects due to ray bounces between the spacecraft surfaces. In addition, complex models for the spacecraft thermal radiation, Bennu Albedo (ALB) and HGA antenna transmission pressure were developed and used (Berry, et al., 2015; Geeraert, et al., 2020). A successful touch-and-go (TAG) maneuver, to collect a sample from Bennu's surface, required predicted state error resolution of tens of meters, and perturbing acceleration characterization at the level of 3 nm/s$^2$ (Antreasian, et al., 2016). This requirement was ultimately met with the implementation of complex models, daily 5 to 8 hour DSN radiometric tracking, Bennu landmark tracking, and altimetry measurements (Getzandanner, et al., 2016).

We have developed a simulation of OREx mission navigation in the vicinity of Bennu, with the added benefit of onboard SLI-based accelerometers. The simulation is performed using the Jet Propulsion Laboratory's (JPL) Mission Analysis, Operations, and Navigation Toolkit (MONTE) software. MONTE allows high fidelity orbit propagation and estimation modeling for virtually any mission scenario (NASA: Jet Propulsion Laboratory, 2021). The simulation uses an OREx "truth" reconstructed trajectory and spacecraft attitude obtained from the Navigation and Ancillary Information Facility (NAIF) database. The NAIF trajectory and attitude are imported directly into the MONTE simulation and the position, velocity, acceleration, and attitude are read off at requested simulation times. The filter nominal trajectory is generated using MONTE's astrodynamic model libraries, populated with relevant spacecraft and relevant celestial body parameters produced by the navigation team to support the mission. The true and nominal trajectories differ based on the fidelity of the non-gravitational force models implemented for the truth reconstruction by the OREx team versus the nominal propagation in MONTE. In addition, the nominal trajectory is modified to include initial condition and gravity errors for the Monte Carlo analysis. The rest of this section details set-up for this OREx simulation.

**3.1  NAIF Truth Trajectory**

Two timeframes were selected for simulation, shown in Table 2, which occur during the Orbit A and B phases of the OREx mission. The Orbit A phase began on December 31, 2018, with a nominal orbit radius of 1.5 km. Orbit A was used to establish Bennu landmark tracking and characterize orbit navigation and maneuver performance. Orbit B began on June 12, 2019, with a nominal radius of 1 km. Orbit B was used primarily for radio science, establishing an accurate Bennu shape model, and determining a feasible landing site for sample collection (Lauretta, et al., 2017).

Table 2: Simulation Timeframes

| Orbit A | Orbit B |
|---|---|
| February 8 – 12, 2019 | July 12 – 16, 2019 |

These two phases were chosen to provide orbits with different radii and Sun-Bennu distances, resulting in distinct gravity and non-gravitational force acceleration profiles and magnitudes. From Orbit A to Orbit B the reduction in altitude results in an order of magnitude increase in perturbing accelerations from spherical harmonics (SPH); while a larger Sun-Bennu distance reduces the SRP and spacecraft thermal radiation magnitudes, shown later in Figure 3. Figure 1 shows the two orbit orientations, indicating the radial (green), transverse (blue), and orbit normal (red) (RTN) directions at the start of each phase. Each orbit is shown from the perspective of Earth, looking directly towards Bennu. Both orbits lie in the terminator plane, with the normal direction aligned with the Sun-vector. The OREx team chose this orientation to remove the complication of OREx passing into Bennu's shadow, thereby providing better stability with respect to SRP and temperature effects (Lauretta, et al., 2017).

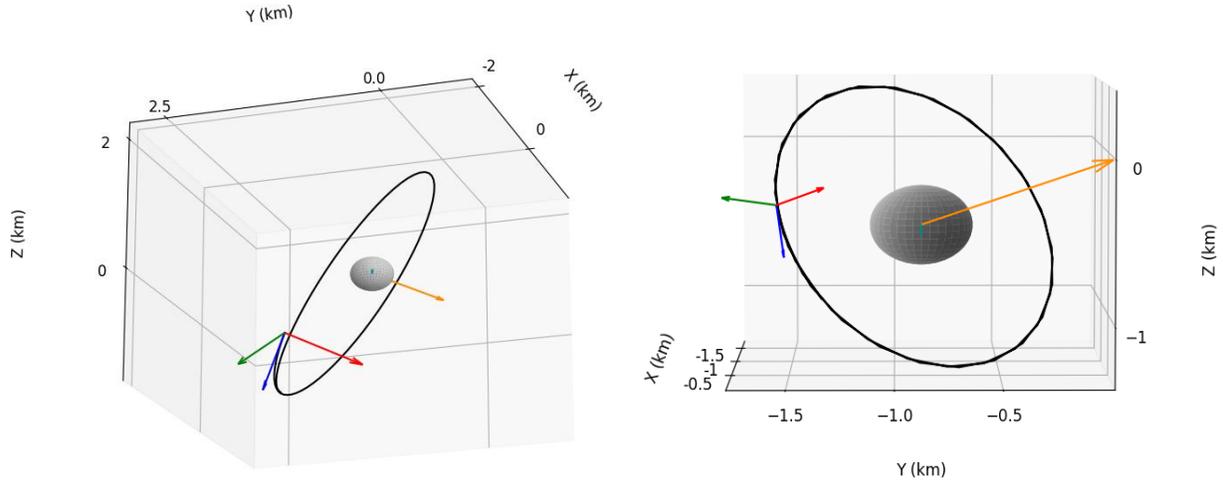

Figure 1: OREx Orbit A (left) and B (right) orbit orientations. Unit vector to Sun shown in gold and viewpoint oriented along unit vector to Earth, shown in cyan. Size of orbit and average asteroid radius shown to scale.

The spacecraft frame is shown in Figure 2. The +X axis is oriented through the face containing the HGA. The +Z axis is oriented along the face containing the low-gain antenna and the +Y axis completes the right-handed frame (Semenov, 2014; Lauretta, et al., 2017). During Orbits A and B, OREx was primarily oriented with Z-axis aligned in the nadir-direction and the +X-axis, aligned with the Sun-vector (hence predominantly in the orbit normal direction). Once a day the spacecraft is slewed to point the HGA antenna towards Earth for DSN radiometric tracking for up to 5 hrs per day (Williams, et al., 2018).

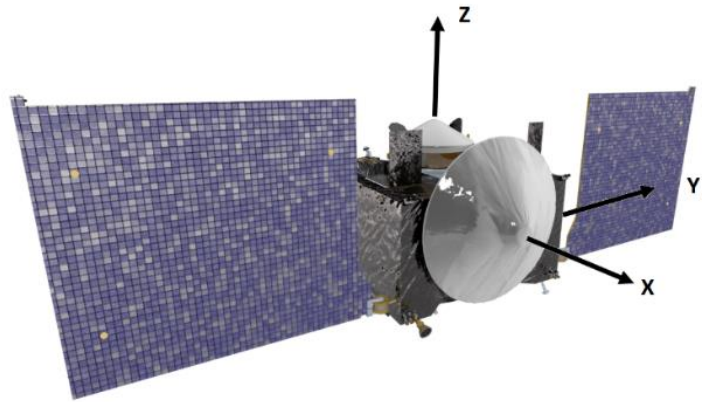

Figure 2: OREx (OSIRIS-REx Asteroid Sample Return Mission, 2022) with spacecraft coordinate axes added (Lauretta, et al., 2017)

For the purposes of this work, the state vectors are propagated and estimated in the filter in the EME2000 frame. The non-gravitational acceleration profiles are presented in the spacecraft XYZ frame due to their strong correlation with the spacecraft pointing directions. The filter results are mapped to the RTN frame to allow for comparison with mission analysis presented by the OREx navigation team.

## 3.2 MONTE Simulated Nominal Trajectory

The nominal trajectory for filter propagation is simulated using point mass (Gm) and spherical harmonic (SPH) gravity, SRP, and Albedo (ALB) acceleration models, populated with relevant OREx and Bennu parameters. A time history of the acceleration due to spacecraft thermal radiation pressure, provided by the NASA/KinetX team, is included in the model validation (J. Leonard and J. Geeraert personal communication, Sept 3, 2021). The set-up for each acceleration model is validated and then used to generate a nominal reference for the filter, populated with the model fidelity available during the mission phase.

### 3.2.1 Acceleration Modeling for OREx

The OREx spacecraft is modelled using nine flat plates to represent the -X and +/-YZ bus faces and the front and back of each solar array; and a parabolic dish for the HGA. The specular and diffuse coefficients, shown in Table 3, are based on Germanium Kapton MLI for the bus and HGA, and general material properties for solar panels, as presented in Kenneally (2019).

Table 3: Spacecraft SRP Model Parameters (Kenneally, 2019)

| Component | NAIF Frame ID | Normal Vector | Spec Ref Coef. | Diff Coeff. | Shape | Area ($m^2$) |
|---|---|---|---|---|---|---|
| +X bus | -64000 | [1, 0, 0] | 0.408 | 0.102 | Parabolic Dish | 2.1 m diameter, 0.4 m depth |
| -X bus | | [-1, 0, 0] | | | Flat Plate | 4.8 |
| +/- Y bus | | [0, +/-1, 0] | | | | 5.76 |
| +/- Z bus | | [0, 0, +/-1] | | | | |
| +/- Y Front Solar Panel | -64012 | [0, 0, 1] | 0.088 | 0.022 | | 4.25 |

| +/- Y Back Solar Panel | -64022 | [0, 0, -1] | 0.00 | 0.05 | | |

The spacecraft bus attitude is provided from the NAIF database. The solar array baseplate frames, fixed to the bus attitude, are then obtained via a 180° rotation about the Z axis and –90° degree rotation about the X axis for the +Y array, and a 90° rotation about X for the -Y array. The rotations of the gimbal frames from NAIF, defined about the baseplate frames, are then used to define the attitude profiles for each solar array (Semenov, 2014).

The filter nominal SRP and ALB profile generation uses the spacecraft shape model and orientation data to determine the force incident on each spacecraft component. The effect on each component is then combined into an overall acceleration. The thermophysical coefficients for Bennu used in the ALB simulation are shown in Table 4.

Table 4: Bennu Thermophysical Properties (Rozitis, et al., 2020)

| Component | Coefficient |
|---|---|
| Albedo | 0.02 |
| Thermal Emissivity | 0.95 |

Both models use the orientation of the spacecraft components to determine effective area, but do not account for self-shadowing i.e., if one surface is blocking another from the incident rays. Bennu's gravity model is simulated with the Gm gravitational parameter and 16x16 SPH field used during the actual NAIF trajectory reconstruction, provided by the NASA/KinetX team (J. Leonard and J. Geeraert personal communication, Sept 3, 2021).

### 3.2.2 MONTE Model Set-Up Validation

To validate each acceleration component set-up in MONTE, the component of interest is evaluated along the NAIF-truth trajectory, populated with the OREx and Bennu parameter values used at the time of NAIF trajectory reconstruction. The profile is then compared to the full

NAIF-truth acceleration minus the profiles of all but the acceleration of interest, each evaluated along the NAIF-truth. This comparison confirms that the correct set-up parameters have been chosen. The full 16x16 SPH field, perfect Gm, and the thermal radiation acceleration profile are used for the validation process. Figure 3 shows the resulting acceleration profiles obtained for the total trajectory (green), Gm (black), SPH (blue), SRP (red), ALB (cyan), and spacecraft thermal radiation (gold) accelerations, evaluated along the NAIF-true trajectory. The acceleration profiles and magnitudes are consistent with those generated by the OREx team for both orbit phases, as presented in Antreasian, et al. (2016) and Leonard, et al. (2019). To determine the difference between the total NAIF reconstructed and combined acceleration models, the Gm, SPH, SRP, ALB, and thermal acceleration models (ScTherm), propagated along the true NAIF trajectory, are subtracted from the total NAIF trajectory acceleration profile, shown in pink on Figure 3.

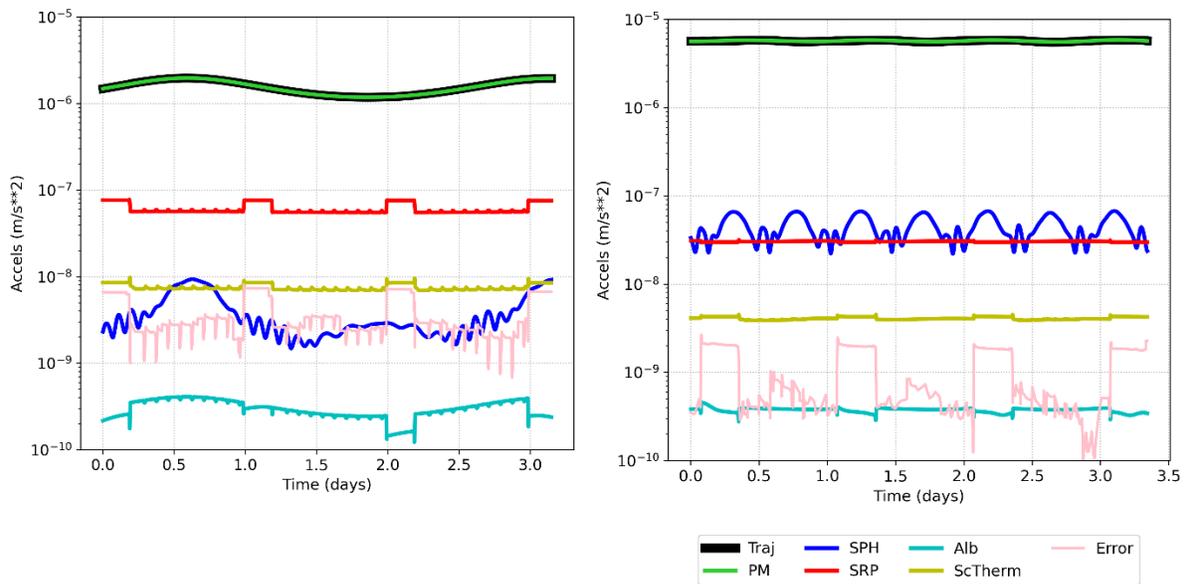

Figure 3: Simulated Acceleration Magnitudes of OREx in Orbit A (left) and B (right)

Figure 4 shows a comparison between the NAIF-true trajectory acceleration minus the Gm and SPH models propagated along the NAIF-true trajectory and the SRP, ALB and thermal radiation models propagated along the NAIF-true trajectory. This shows the difference between the NAIF-true non-gravitational acceleration, used as the truth for this study, and non-gravitational acceleration models set-up in MONTE. This is the pink profile from Figure 3, shown in terms of its XYZ components in the spacecraft frame in Figure 4.

Figure 5 shows the differences between the two acceleration profiles presented in Figure 4. The difference between the true and modelled accelerations evaluated along the true trajectory, is primarily due to the 9 plates plus cone HGA SRP model used. This SRP model uses imperfect spectral and diffusion coefficients, an overly simplistic HGA model, and does not account for self-shadowing or solar ray bounces between surfaces. The mismodeling is made more apparent by once per day attitude maneuvers to orient the high gain antenna (HGA) toward the Earth. The change of Sun exposure on the different spacecraft surfaces in the Earth pointing, versus the nominal Sun pointing orientations, results in significant jumps in the SRP profile, that can be seen at the start of each day in Figure 5. Additional spikes are also present in the Orbit A total and differenced profiles when ORX performs imaging sweeps of Bennu for optical navigation, seen between the larger daily HGA-pointing maneuvers (Williams, et al., 2018).

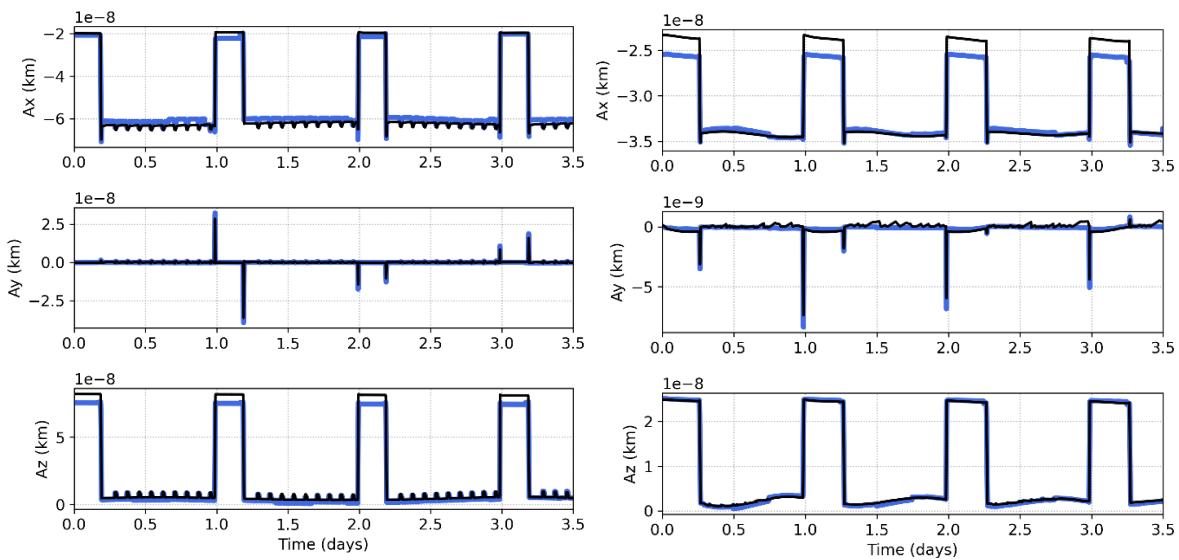

Figure 4: Comparison of the NAIF-true non-gravitational profile (blue) and the non-gravitational acceleration models (SRP, ALB, and thermal radiation) propagated along the NAIF-true trajectory (black) for Orbit A (left) and Orbit B (right) expressed in the spacecraft frame

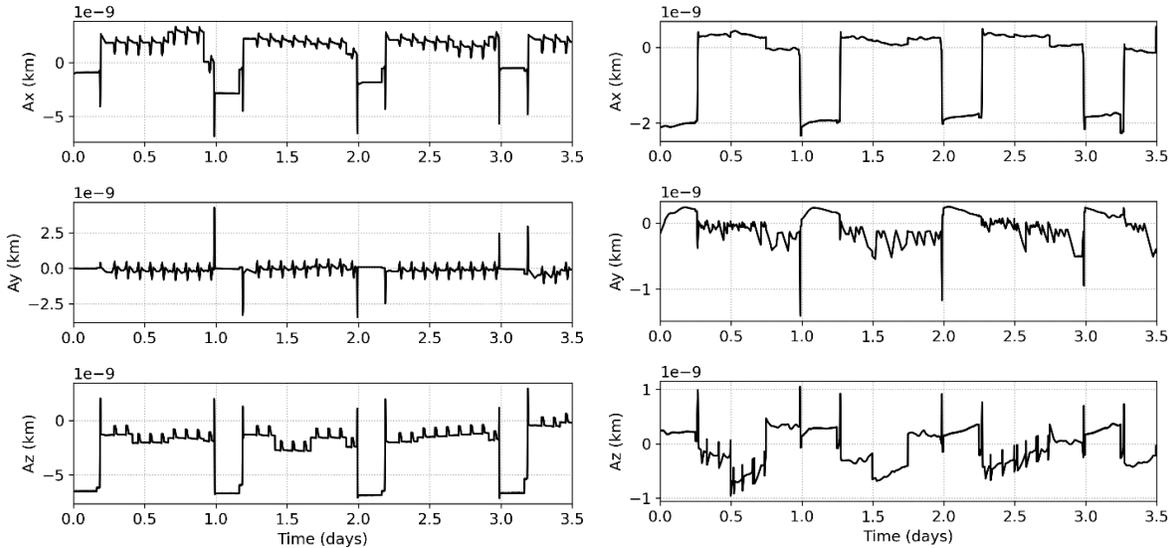

Figure 5: Difference between the NAIF-true non-gravitational profile and the non-gravitational acceleration models (SRP, ALB, and thermal radiation) propagated along the NAIF-true trajectory for Orbit A (left) and Orbit B (right) expressed in the spacecraft frame

The differences introduced by the SRP mismodeling are consistent with those documented by the OREx team for Orbit A when using a simplified SRP model (Geeraert, et al., 2020). The team saw differences on the order of $1 \times 10^{-9}$ m/s$^2$ in X and $6 \times 10^{-9}$ m/s$^2$ in Z in the Earth pointing orientation. To reduce these effects, they developed and incorporated complex HGA and ray-trace SRP models into the OREx mission navigation process (Leonard, et al., 2019). For our study, onboard SLI-based accelerometers will measure the total non-gravitational accelerations, including the signature that is mismodeled in the filter nominal propagation, seen in Figure 5. This will allow an analysis of how well the mismodeled accelerations can be compensated for in the filter using the accelerometer measurements. If the mismodeling can be sufficiently captured by the accelerometer measurements, then modeling requirements, such as the need for the complex SRP model, could potentially be reduced.

# 4 OREx Measurement Simulation and Navigation Filter Set-Up

The position and velocity of OREx are estimated in a conventional Kalman filter (CKF). DSN radiometric tracking measurements and SLI-based accelerometer measurements are simulated along the true trajectory, including relevant sensitivity levels. The filter propagates the nominal trajectory and processes the DSN radiometric and accelerometer measurements. The initial condition error for Orbit A is 1 m position and 0.1 mm/s velocity 1σ. This error is based on an approximation of the trajectory error predicted by the OREx team at Data Cutoff (DCO) in Reference (Geeraert, et al., 2020), when using an optimized antenna +9-plate SRP model. For Orbit B, initial condition errors of 0.3 m and .03 mm/s 1σ are used, based on the state errors predicted by the OREx team at DCO in Reference (Williams, et al., 2018), with and without desaturation maneuvers. The nominal trajectory includes Gm and 3x3 SPH gravity, with 1σ gravitational parameter error (Gm) of $2.44 \times 10^{-12}$ km$^3$/s$^2$ (.15% of Gm) (Williams, et al., 2018) (Antreasian, et al., 2016; Williams, et al., 2018; Leonard, et al., 2019; French, 2020). The simple 9-plate + HGA spacecraft model is used for SRP and ALB propagation. During navigation analysis of the OREx mission, presented in Leonard, et al. (2019) and Berry, et al. (2015), a 3σ SRP scale factor error of 10% was used. For the purposes of this simulation, SRP errors are present due to the simplified SRP model, so no additional scale factor error is applied. Acceleration due to the spacecraft thermal radiation pressure is not included in the nominal model in the filter.

## 4.1 Simulated DSN Measurements

During both Orbit A and B phases of OREx radiometric tracking from the DSN was available for approximately 5 hrs per day on the HGA with up to an additional 3 hrs on the low-gain antenna. For the actual OREx mission, the two-way range noise was weighted at 3 m 1σ to account for hardware biases, ionosphere, and solar plasma effects. The X-band (~8.45 GHz) Doppler noise for a 60 s count time was weighted at 0.1 mm/s 1σ, 0.00563 Hz 2-way (Williams, et al., 2018). In our simulation, two-way range and Doppler measurements are created at a 60 s time step, including the 3 m and 0.1 mm/s noise error respectively. The simulated measurement passes are aligned with the HGA Earth-pointing attitude times, i.e., the times when the daily large steps occur in the profiles shown in Figure 4, using the DSN station in view during that interval. These passes occur daily and are approximately 5 hrs long.

### 4.2 Simulated Accelerometer Measurements

Accelerometers sense the non-gravitational forces incident on a spacecraft (Titterton & Weston, 2004; Jekeli, 2005; Jekeli, 2006). To simulate accelerometer measurements of the OREx true trajectory, the validated Gm and 16x16 SPH gravity model is evaluated along the trajectory. This gravitational acceleration ($\vec{a}_{true, Grav}$) is then subtracted from the total acceleration profile ($\vec{a}_{true, Total}$), leaving the non-gravitational acceleration ($\vec{a}_{NG}$), as shown in Eq. (2).

$$\vec{a}_{NG} = \vec{a}_{true, Total} - \vec{a}_{true, Grav} \qquad (2)$$

Following the work of Ely, Heyne, and Riedel (2012), the accelerometer measurements are simulated as delta-velocity measurements of Euler integrated accelerations over the measurement sample period, shown in Eq. (3),

$$\Delta V(t_n) = \sum_{i=1}^{n-1} [I + \Sigma + \Lambda + \Omega][a_{ng}(t_i) + b + v(t_i)][t_{i+1} - t_i] \qquad (3)$$

where measurements are taken every 60 s and $n$ is the sample counter, $v$ is the Gaussian white noise accelerometer error, $\Sigma$ is the scale factor error, $\Lambda$ is the misalignment error, $\Omega$ is the orthogonality error, and $b$ is the bias error. Internally, the accelerometer is accumulating the acceleration over a 60 s sampling interval, at an internal frequency rate.

Based on the range of potential performance metrics for SLI-based accelerometers, multiple accelerometer sensitivities are investigated, shown in Table 5. The sensitivities are presented as amplitude spectral density power and per shot following the calculation shown in Eq. (1).

Table 5: Investigated Accelerometer Sensitivities

| Sensitivity per Shot | $1\times10^{-8}$ m/s$^2$ /shot | $1\times10^{-9}$ m/s$^2$ /shot | $1\times10^{-10}$ m/s$^2$ /shot |
|---|---|---|---|
| Amplitude Spectral Density | $7.75\times10^{-8}$ m/s$^2$ /$\sqrt{\text{Hz}}$ | $7.75\times10^{-9}$ m/s$^2$ /$\sqrt{\text{Hz}}$ | $7.75\times10^{-10}$ m/s$^2$ /$\sqrt{\text{Hz}}$ |

Figure 6 shows the simulated accelerometer measurements, observed along the true trajectory, and the measurements computed by the filter along the nominal trajectory. The computed measurements, propagated along the nominal trajectory, include a realization of the injected error sources and initial condition error. The observed measurements shown are generated for the $1\times10^{-9}$ m/s$^2$ accelerometer sensitivity.

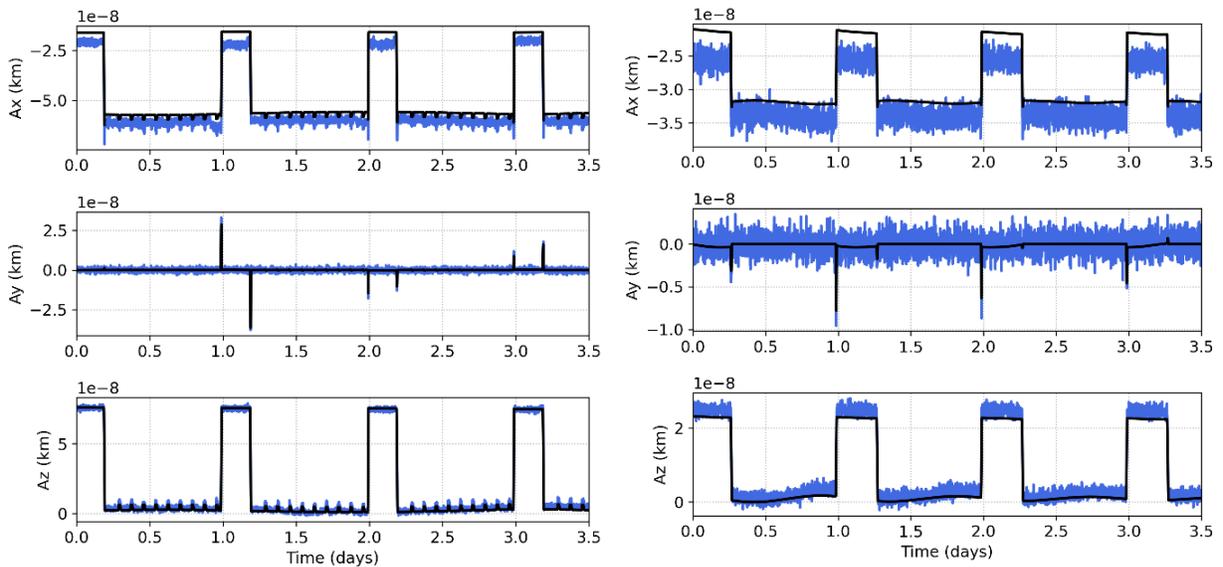

Figure 6: XYZ components of the observations simulated along the NAIF-true trajectory (blue) and computed (black) measurements simulated along the filter nominal, for a $1\times10^{-09}$ m/s$^2$ sensitivity, for Orbit A (left) and Orbit B (right) expressed in the spacecraft frame

The discrepancy between the simulated observed and computed measurements is due to the simplified models used in the filter nominal propagation and the initial condition errors applied. The most significant effect is from the simplified model of the HGA on the +X spacecraft face, which appears as discrepancy in the X and Z directions, depending on whether the spacecraft is in the Earth or Sun-pointing orientation. In addition, the thermal radiation pressure effect is greatest in the X spacecraft direction. Thus, its exclusion from the nominal propagation leads to a discrepancy between the observed and computed measurements in that direction.

Figure 7 shows the difference between the observed measurements, simulated along the NAIF-true trajectory, and the computed measurements, simulated along the filter nominal, for both orbits. The accelerometer sensitivity of $1 \times 10^{-9}$ m/s$^2$ is sufficient to observe the overall features of the non-gravitational force modeling error, which are on the order of $0.5\text{-}1.5 \times 10^{-8}$ m/s$^2$ for Orbit A and $2.5\text{-}5 \times 10^{-9}$ m/s$^2$ for Orbit B. However, some of the smaller features are obscured by the measurement noise.

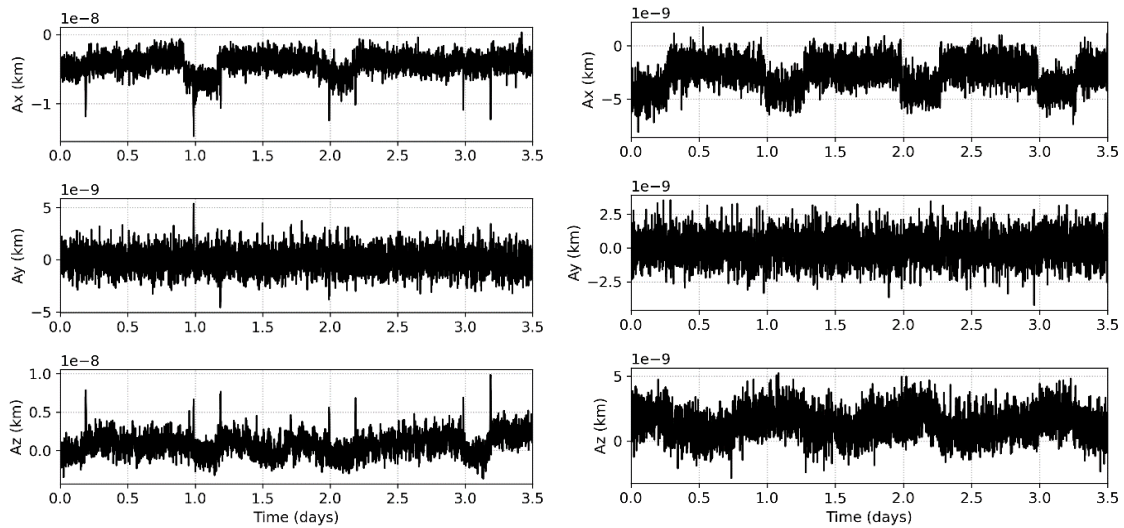

Figure 7: XYZ components of the observations simulated along the NAIF-true trajectory minus computed measurements simulated along the filter nominal, for a $1 \times 10^{-09}$ m/s$^2$ sensitivity, for Orbit A (left) and Orbit B (right) expressed in the spacecraft frame

## 4.3 Filter and Accelerometer Measurement Processing Set-Up

The filter state ($X$) includes the XYZ position and velocity deviations from the nominal for the OREx spacecraft in the EME2000 frame. In addition to the position and velocity states the filter includes stochastic acceleration states, estimated to account for the unmodeled accelerations. The stochastic acceleration state for each component is modeled as a mean-zero white noise process, in the spacecraft frame. The stochastic acceleration state ($a_i$) is incorporated into the filter as shown in Eq. (4) (shown for the X axis in the spacecraft frame),

$$\begin{bmatrix} x_{x,i+1} \\ \dot{x}_{x,i+1} \\ a_{x,i+1} \end{bmatrix} = \begin{bmatrix} 1 & \Delta t & \Delta t^2/2 \\ 0 & 1 & \Delta t \\ 0 & 0 & 0 \end{bmatrix} \begin{bmatrix} x_{x,i} \\ \dot{x}_{x,i} \\ a_{x,i} \end{bmatrix} + \begin{bmatrix} 0 \\ 0 \\ 1 \end{bmatrix} w_{x,i} \tag{4}$$

where $\Delta t$ is the filter step-size, $\dot{x}_i$ is the velocity state, $x_i$ is the position state, and $w_{x,i}$ is the white noise process (Ely, Heyne, & Riedel, 2012). This allows the filter to absorb unmodeled accelerations into the stochastic XYZ state informed by the noise sigma of $w_{x,i}$. The noise strength is determined by the size of the unmodeled gravitational and non-gravitational force acceleration errors induced by the mismodeling in the nominal trajectory.

The accelerometer measurements relate to the stochastic acceleration state via the measurement equation (shown for the X-axis in the spacecraft frame),

$$\Delta V_{x,i} = a_{x,i} \Delta t + v_{x,i} \tag{5}$$

and subsequent measurement partial matrix,

$$\frac{\partial \Delta V_{x,i}}{\partial X_i} = [0_{1x3} \quad 0_{1x3} \quad \Delta t \quad 0 \quad 0]. \tag{6}$$

The accelerometer measurements are processed directly in the filter as done by Ely, Heyne, and Riedel (2012). This is in contrast to a dead reckoning approach commonly used in inertial navigation systems, where accelerometer measurements are used to propagate the nominal trajectory, as presented in Jekeli (2005).

# 5 OREx Simulation Results

The filter was run for full and reduced DSN pass cases for Orbit A and B. The first scenario uses all the daily passes, representing the pass configuration available during the actual OREx mission (Antreasian, et al., 2016). The second scenario removes the third and fourth pass for Orbit A and B respectively. Ten unique profiles of initial condition and measurement error were constructed. These profiles were applied to each scenario for a DSN only case and for several accelerometer sensitivity configurations. Each of the ten run sets used the same initial condition error and DSN and accelerometer measurement sensitivity realizations. The average position RMS for each component for each of the ten runs are shown in Table 6, expressed in the RTN frame. The RMS is calculated from the start of the 3rd DSN pass for Orbit A and 2nd pass for Orbit B, to allow for filter convergence. Figure 8 shows the position estimation error and 1σ bounds for the daily pass (passes shown in grey) configuration, for DSN measurements only (blue) and $1 \times 10^{-9}$ m/s$^2$ accelerometer measurements (magenta) configurations.

Table 6: Orbit A and B RMS

| Orbit | Accelerometer Sensitivity per Shot (m/s$^2$) | RMS (m) | | | | | |
|---|---|---|---|---|---|---|---|
| | | Full DSN | | | Reduced DSN | | |
| | | R | T | N | R | T | N |
| A | None | 5.0 | 11.8 | 2.4 | 5.1 | 19.6 | 2.7 |
| | $1 \times 10^{-08}$ | 3.0 | 6.1 | 1.8 | 3.3 | 8.4 | 2.1 |
| | $1 \times 10^{-09}$ | 3.0 | 5.9 | 1.8 | 3.2 | 8.2 | 2.0 |
| | $1 \times 10^{-10}$ | 2.9 | 5.9 | 1.8 | 3.2 | 8.1 | 2.0 |
| B | None | 1.3 | 4.8 | 0.5 | 1.3 | 6.9 | 0.7 |
| | $1 \times 10^{-08}$ | 0.8 | 3.4 | 0.4 | 0.9 | 3.6 | 0.4 |
| | $1 \times 10^{-09}$ | 0.8 | 2.6 | 0.4 | 0.9 | 3.5 | 0.4 |
| | $1 \times 10^{-10}$ | 0.8 | 2.6 | 0.4 | 0.9 | 3.4 | 0.4 |

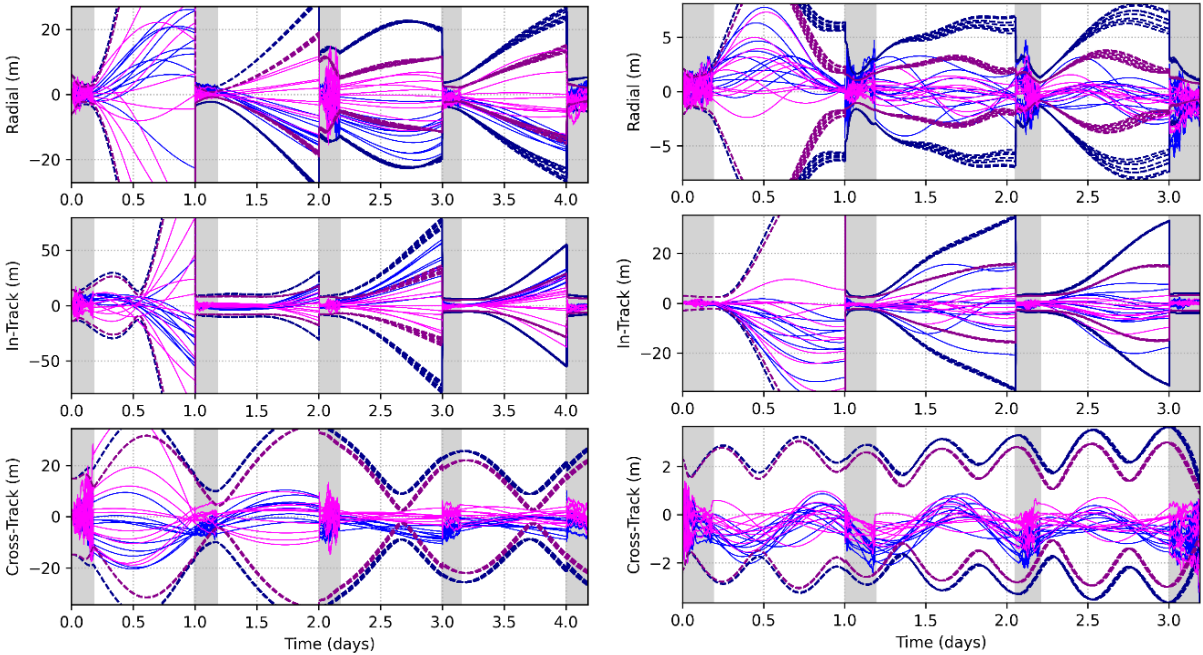

Figure 8: Daily DSN passes – position errors and 1σ covariance bounds (dark/dashed) without (blue) and with (magenta) $1 \times 10^{-9}$ m/s$^2$ sensitivity accelerometer measurements for Orbit A (left) and Orbit B (right) – RTN frame

The inclusion of accelerometer measurements improves the navigation performance for both orbit phases. Using accelerometers with sensitivity less than $1 \times 10^{-9}$ m/s$^2$ results in only marginal improvement. Examining Figure 6 and Figure 7, the primary benefit of the accelerometers is from capturing the non-gravitational force profile features that are on the order of $1 \times 10^{-09}$ m/s$^2$. During the real Orbit A mission phase, SRP scale factor and gravity parameters were estimated, allowing position resolution to 10-meter levels (Antreasian, et al., 2016; Williams, et al., 2018). SRP is the largest acceleration for Orbit A (aside from Gm), and an order of magnitude larger than SPH, so capturing the non-gravitational forces with the accelerometers allows for an improvement in estimation from tens of meters to less than 10 m. For Orbit B, the SRP surface forces and the uncertainty in the SPH gravitational forces are of very similar magnitude. The addition of the accelerometers for Orbit B improves the position resolution by a few meters. The reduction in error overall is less in the case of Orbit B than A, but there is a relative reduction, from the DSN only to the $1\times10^{-9}$ m/s$^2$ sensitivity case, of 40% in radial, 50% in in-track and 20% in cross-track errors, for both orbits.

Overall, we find that the navigation performance achieved in both orbit scenarios, with reduced DSN tracking and onboard accelerometers, is comparable to or better than the use of a full DSN daily tracking schedule with no accelerometers. Using the onboard accelerometers helps resolve the non-gravitational accelerations, subsequently improving the ability to determine the gravity field from the radiometric measurements. Furthermore, the use of SLI-based accelerometers could reduce the dependence on high accuracy non-gravitational force modeling and expanded filter state vectors to estimate systematic disturbances, allowing for the navigation filter state vector to includes only the vector components of position, velocity, and stochastic acceleration.

## 6 Conclusion

The advancement of compact precision accelerometers based on shaken lattice interferometry (SLI) provides an opportunity to consider future space exploration missions with greater autonomy and flexibility to observe and adapt to unexpected dynamical conditions. In this study, we analyzed of the use of precise SLI-based accelerometers for an OREx type mission, choosing this mission because of the unique challenges faced in navigating at very low altitudes above a very small body. These small orbits, only weakly held by Bennu's gravity, are substantially perturbed by solar radiation pressure – which is highly variable due to eclipse conditions and significant changes in spacecraft orientation. It is therefore an ideal case study to establish the levels of accelerometer performance useful for future space missions. For this scenario we found that a sensitivity of $1\times10^{-9}$ m/s$^2$ is required to capture relevant surface accelerations and have a meaningful impact on navigation performance. Reduction in acceleration sensitivity below that level provides no appreciable navigation benefit without a corresponding improvement in knowledge of the gravitational field. In addition, the utilization of onboard accelerometers was shown to allow for a reduction of at least a day in required DSN tracking while maintaining similar navigation performance. This study used real reconstructed truth from the NAIF database to provide the most realistic simulation of dynamical errors possible, and JPL's MONTE software to provide a high-fidelity simulation comparison.

The analysis presented did not consider the potential for estimating Bennu's gravity field using a combination of DSN and SLI-based accelerometer measurements. This configuration would be similar to the high-fidelity Earth-gravity mapping performed by the GRACE and GOCE missions (Christophe B. , 2013). The precise characterization of the surface accelerations would reduce the need to estimate parameters of complex models for SRP and thermal radiation. This would eliminate the aliasing of non-gravitational accelerations into the Gm and SPH estimation and facilitate better resolution of the gravity field. Migliaccio et al. (2019) proposed the Mass Observation with Cold Atom Sensors in Space (MOCASS) mission, a GRACE/GOCE analog gravity mapping mission using onboard cold atom sensors, but to date it remains only a mission concept (Pivetta, Braitenberg, & Barbolla, 2021).

## Acknowledgements

The authors are grateful to Dr. Jason Leonard and Dr. Jeroen Geeraert from KinetX for providing data and guidance for the use of the OREx NAIF reconstructed trajectories. Additional thanks to Liang-Ying Chih and Professor Murray Holland for the helpful discussions on the operation of CAI/SLI-based accelerometers. The work by M. Rybak and P. Axelrad is supported by the NSF QII–TAQS award number 1936303.